\documentclass{article}
\usepackage{spconf,amsmath,graphicx,booktabs,multirow,color,xspace}

\usepackage[breaklinks=true,bookmarks=false]{hyperref}

\def\cifar{CIFAR10\xspace}
\def\mnist{MNIST\xspace}
\def\imagenet{ImageNet\xspace}
\def\lfw{LFW\xspace}
\def\pascal{PASCAL-VOC12\xspace}
\def\cnn{CNN\xspace}
\def\cnns{CNNs\xspace}
\def\relu{ReLU\xspace} 
\def\bpp{bpp\xspace}
\def\psnr{PSNR\xspace}
\def\psnrv#1{#1 db\xspace} 
\def\ssim{SSIM\xspace}

\def\figref#1{Figure~\ref{figure:#1}}
\def\tabref#1{Table~\ref{table:#1}}
\def\secref#1{Sec.~\ref{sec:#1}}

\def\eqref#1{Eq.~(\ref{eq:#1})}

\def\ie{\emph{i.e.}\xspace}

\def\cf{\emph{c.f.}\xspace}

\def\etc{\emph{etc.}\xspace}
\def\etal{\emph{et al.}\xspace}

\title{End-to-End Trained \cnn Encoder-Decoder Networks For Image Steganography}
 \name{Atique ur Rehman, Rafia Rahim, Shahroz Nadeem, Sibt ul Hussain\footnote{Email:first.lastname@nu.edu.pk}
 }
 \address{Reveal.ai (Recognition, Vision \& Learning) Lab\\National University of Computer \& Emerging Sciences (NUCES-FAST), Islamabad, Pakistan.}

\begin{document}
%
\maketitle

\abstract{All the existing image steganography methods use manually crafted features to hide binary payloads into cover images. This leads to small payload capacity and image distortion. Here we propose a convolutional neural network based encoder-decoder architecture for embedding of images as payload. To this end, we make following three major contributions: (i) we propose a deep learning based generic encoder-decoder architecture for image steganography; (ii) we introduce a new loss function that ensures joint end-to-end training of encoder-decoder networks; (iii) we perform extensive empirical evaluation of proposed architecture on a range of challenging publicly available datasets (\mnist, \cifar, \pascal, \imagenet, \lfw) and report \emph{state-of-the-art} payload  capacity at high \psnr and \ssim values.
}\\
\section{Introduction}
In the field of information security steganography and steganalysis are considered as two important techniques~\cite{li2011survey, hussain2013survey, subhedar2014current}. 
Steganography is used to conceal secret information (\ie a message, a picture or a sound) also known as payload into another non-secret object (that can be an image, a sound or a text message) also known as cover object, such that both the secret message as well as its content remain invisible. Thus in steganography, the overall goal is to conceal a payload into a cover object without compromising the fidelity of cover object and payload. Media files (such as sound and image) are preferred choices for cover objects because of their size, where images have recently  gained a lot of popularity as cover objects in research community~\cite{zeng2016large, li2011survey, subhedar2014current}.  Steganalysis serves as adversarial to steganography and targets to identify cover objects and their payloads.

In image steganography, most of the work has been done to hide a specific text message into a cover image. Thus the focus of all the existing techniques has been finding either noisy regions or low-level image features such as edges, textures, \etc, in cover image for embedding maximum amount of secret information without distorting the original image. 

For instance,  Least Significant Bit (LSB) substitution methods have been extremely popular for image steganography~\cite{zeng2016large} due to their simplicity.  These LSB methods replace the least significant bit of a chosen pixel of cover object by a single bit information from a secret message. However, since the pixels of cover object are updated in a disjoint manner,  this leads to image distortion due to change in distribution of LSBs of image pixels. Thus, it becomes easy to detect both the cover object and its payload. Recently, researchers have proposed range of improvements on basic LSB approach. For instance, Yang \etal~\cite{yang2009high} use image content  (texture, edges and pixels brightness) to estimate number of LSBs for robust hiding of payload. 

Alternatively, other researchers have used domain knowledge to extract other primitive features for information hiding. For example, Holu \etal~\cite{holub2012designing} use weights obtained from Wavelets in their WOW (Wavelet Obtained Weights) algorithm for finding noisy and textured regions of image for embedding the secret message. Zhao \etal and Tai \etal~\cite{zhao2011reversible, tai2009reversible} propose modifications in image histogram representations as a data hiding technique.  Saif \etal~\cite{islam2014edge} use cover image edges for extending the payload embedding capacity and avoiding the detection during steganalysis.

In short, all these existing works use domain knowledge to identify basic image features for binary payload hiding while being robust to distortion in cover image appearance.
\begin{figure*}[!ht]
\begin{center}
\includegraphics[width=\textwidth]{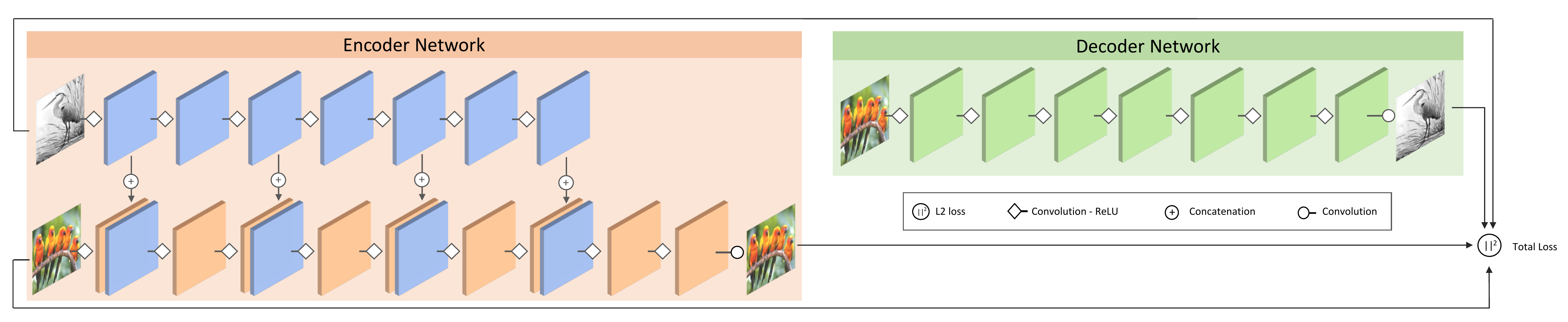}
 \caption{Pictorial representation of encoder and decoder networks architecture. Top row in encoder network represents the guest branch while bottom row represents host branch. }
\label{figure:architecture}  
\end{center}
\end{figure*}
\begin{table*}[!ht]
\vspace*{-1em}
\centering
\resizebox{\textwidth}{!}{%

\begin{tabular}{@{}llclllllllccc@{}}
\multirow{4}{*}{\rotatebox[origin=c]{90}{Encoder}}&
Layers & Input & Conv+ReLU & Conv+ReLU & Conv+ReLU & Conv+ReLU & Conv+ReLU & Conv+ReLU & Conv+ReLU & Conv+ReLU & Conv+ReLU & Conv \\
&Host Branch Filters & $300\times300\times3$ & $3\times3\times16$ & $3\times3\times16$ & $3\times3\times16$ & $3\times3\times16$ & $3\times3\times16$ & $3\times3\times16$ & $3\times3\times16$ & 
\multirow{3}{*}{$1\times1\times16$} & \multirow{3}{*}{$1\times1\times8$} & \multirow{3}{*}{$1\times1\times3$} \\
&Merging &  & Concat &  & Concat &  & Concat &  & Concat &  &  &  \\
&Guest Branch Filters & $300\times300\times1$ & $3\times3\times16$ & $3\times3\times16$ & $3\times3\times16$ & $3\times3\times16$ & $3\times3\times16$ & $3\times3\times16$ & $3\times3\times16$ &  &  & 
\\
\\
\end{tabular}
}%

\resizebox{\textwidth}{!}{\tiny
\begin{tabular}{@{}llcccccccc@{}}
\hline
\\
\multirow{2}{*}{\rotatebox[origin=c]{90}{Decoder}}&
Layers & Input & Conv+ReLU & Conv+ReLU & Conv+ReLU & Conv+ReLU & Conv+ReLU & Conv+ReLU &  Conv \\
&Filters & $300\times300\times1$ & $3\times3\times16$ & $3\times3\times16$ & $3\times3\times8$ & $3\times3\times8$ & $3\times3\times3$ & $3\times3\times3$ & $1\times1\times1$ 
\end{tabular}
}%
\caption{Details of the layers and filters used in encoder and decoder networks.}
\label{table:architecture}
\end{table*}

In comparison to earlier work, here we address a generic problem of hiding a real-world image (payload) into another real-world image (cover). 
According to our best of knowledge no such work exist, that takes both cover and payload directly as images. Note that although here we propose using images for both cover and payload, other information such as  text can be considered as special case of this solution, where text can be converted into a bitmap representation and then can be used as payload with our method. 

Thus to this end, we propose a novel and completely automatic steganography method for hiding one image to another. For this, we design a deep learning network that automatically identifies the  best features from both cover and payload images to merge information.   The biggest advantage of our this approach is that its generic and can be used with any type of images, to validate this we test our approach on variety of publicly available datasets including \imagenet~\cite{russakovsky2015imagenet},  \mnist~\cite{lecun1998mnist}, \cifar~\cite{krizhevsky2014cifar}, \lfw~\cite{huang2008labeled} and  \pascal~\cite{everingham2010pascal}.

\textbf{Overall our main contributions} are as follows: (i) we propose a deep learning based generic encoder-decoder architecture for image steganography; (ii) we design a new loss function that ensures joint end-to-end training of encoder-decoder networks; (iii) we perform extensive empirical evaluation of proposed architecture on range of challenging publicly available datasets and report \emph{state-of-the-art} payload   capacity at high \psnr and \ssim values. Specifically, using our proposed algorithm we can reliably embed a unary channel image ($m\times n$ pixels) into a color image ($m\times n \times 3 $ pixels). Our experiments show that we can achieve this payload of 33\% (on average 8 \bpp) with the average \psnr values of \psnrv{32.9} (SSIM =0.96) for cover and \psnrv{36.6} (SSIM=0.96) for recovered payload image -- \cf \secref{experiments} for further details.\\

%
The rest of the paper is organized as follows. \secref{architecture} gives details on encoder-decoder architecture. \secref{experiments} discusses in detail different datasets, parameter settings and our results. Finally, \secref{conclusion} concludes the paper with relevant discussion.

\section{Methodology}\label{sec:architecture}

We train end-to-end a pair of encoder and decoder Convolutional Neural Networks (\cnns) for creating the hybrid image from pair of input images, and recovering the payload image from input hybrid image -- \cf \figref{architecture} for architecture details. Here, we make use of observation that CNN layers learns a hierarchy of image features from low-level generic to high-level domain specific features. Thus our encoder identifies specific features from cover image to hide the details from the payload images, while decoder learns to separate those hidden features from the ``hybrid'' image.

Specifically, the encoder network takes two images  (\ie a ``host'' cover image and a ``guest'' payload image) as input and produces a single hybrid output image. Thus, the goal of encoder network is to produce a hybrid image, that remains visually identical to the host image but should also contain the guest image content in it. The decoder network takes as input the encoder produced hybrid image and recovers the guest image from it. The goal of decoder network is to recover the guest image from the input hybrid that remains visually similar to input guest image of encoder.

Let $I_h$ and $I_g$ represent input host and guest images to encoder, while $O_e$ and $O_d$ represent the output hybrid image and output decoder image respectively, then the complete loss function for encoder and decoder network can be written as:
\begin{equation}
L(I_g,I_h)=\alpha||I_h-O_e||^2 + \beta ||I_g-O_d||^2 + \lambda (||W_e||^2 + ||W_d||^2) \label{eq:cfun}
\end{equation}

Here $W_e$ and $W_d$ represent the learned weights for the encoder and decoder networks respectively while $\alpha$ and $\beta$ are controlling parameters for encoder and decoder. The first term in loss function defines encoder loss and the second one decoder loss. 

\subsection{Encoder Architecture}


The encoder network at the input end has two parallel branches named as guest branch and host branch. Guest branch receives the input guest image 
$I_g$ and uses a sequence of convolution and \relu layers to decompose the input image into low-level (edges, colors, textures, \etc) and high level features. Host branch receives the input host image $I_h$ and uses a sequence of convolution and \relu layers (except the last layer which does not include \relu layer) to decompose the input image into a hierarchy of feature representations and merge the extracted representations of guest image into host image. 

Precisely, for merging the information from guest image, encoder concatenates the extracted feature maps from each alternating layer of guest branch (starting from input) to the corresponding output features maps of host branch. This procedure is repeated up to a layer of depth $k$ (we found $k=7$ as the best parameter), at this point we completely merge the guest branch features into host branch and guest branch cease to exist. After merging a further sequence of convolution and \relu layers are used before the final convolution layer which produces as output hybrid image $O_e$.

%
%

\subsection{Decoder Architecture}
Our decoder network receives the encoder produced hybrid image $O_e$ as input and runs it through sequence of convolution and \relu layers (except the last layer which does not include \relu) to recover  the concealed representation $O_d$ of guest image $I_g$.


We also experimented with other design choices, however in our initial experiments this architecture comes  out as the best choice. During training both encoder and decoder are trained end-to-end using the joint loss function -- \cf \eqref{cfun}. However during testing both encoder and decoder are used in disjoint manner.

\section{Experiments and Results}\label{sec:experiments}

\begin{figure*}[htb!]
\begin{center}
\resizebox{\textwidth}{!}{
\begin{tabular}{@{}c@{}c@{}c@{}c|c@{}c@{}c@{}c@{}}

\includegraphics{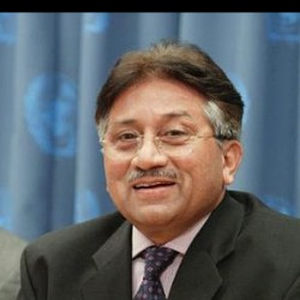}&
\includegraphics{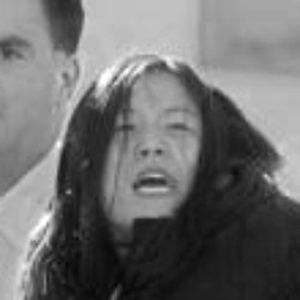}&
\includegraphics{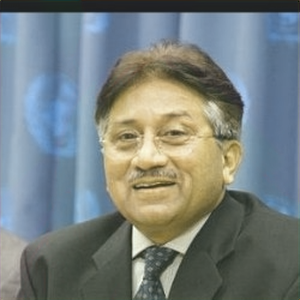}&
\includegraphics{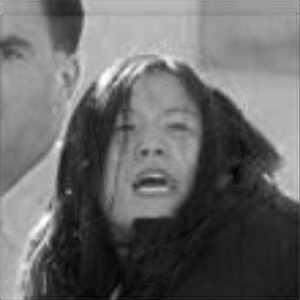}&

\includegraphics{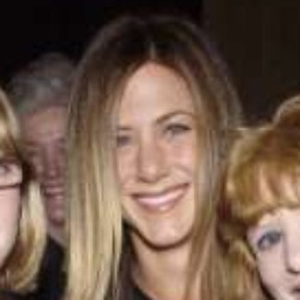}&
\includegraphics{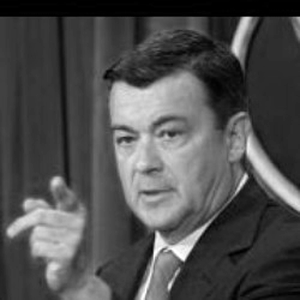}&
\includegraphics{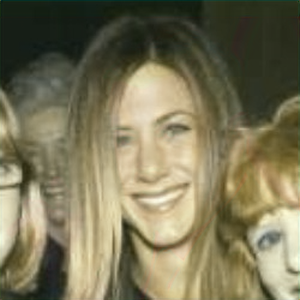}&
\includegraphics{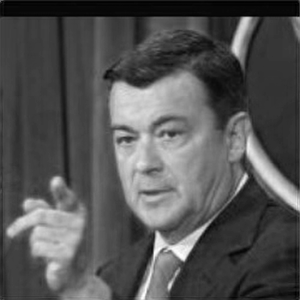} \\

\includegraphics{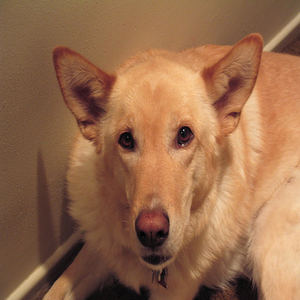}&
\includegraphics{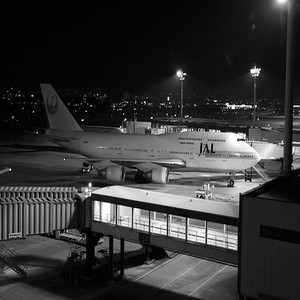}&
\includegraphics{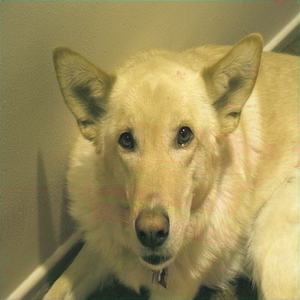}&
\includegraphics{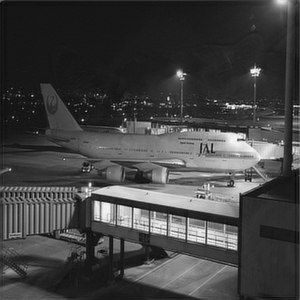}&

\includegraphics{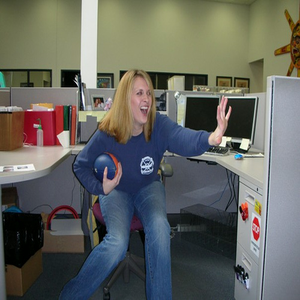}&
\includegraphics{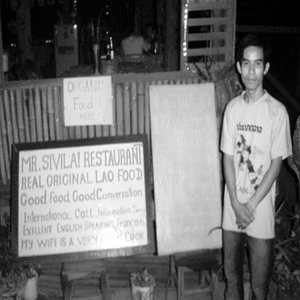}&
\includegraphics{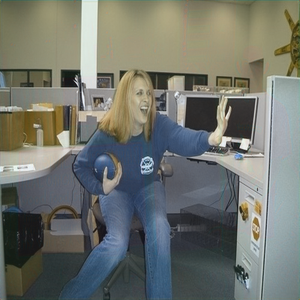}&
\includegraphics{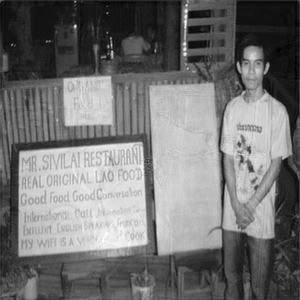} \\

\includegraphics{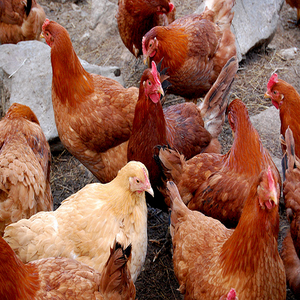}&
\includegraphics{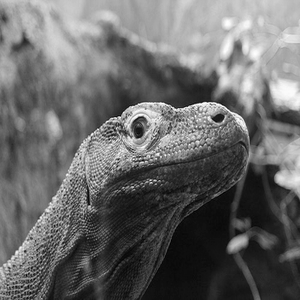}&
\includegraphics{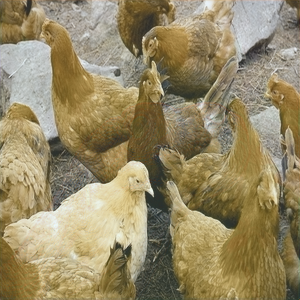}&
\includegraphics{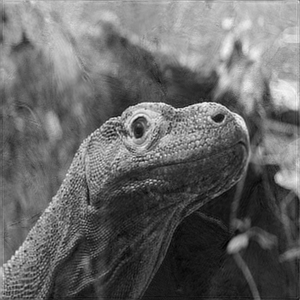}&

\includegraphics{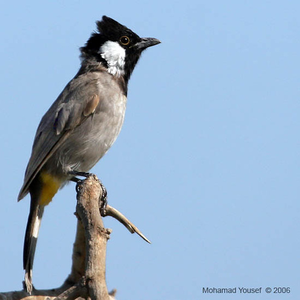}&
\includegraphics{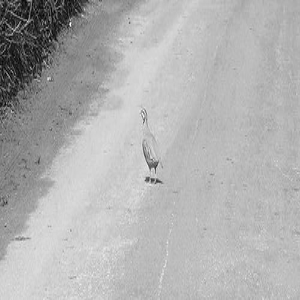}&
\includegraphics{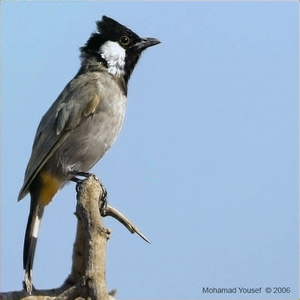}&
\includegraphics{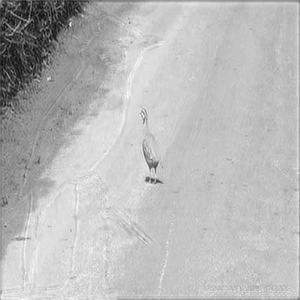} \\
\end{tabular}
}
\end{center}
\caption{Sample results of our algorithm on \lfw (top row), \pascal (middle row) and \imagenet (bottom row) images. In each subfigure, first column represents the cover image $I_h$, second the payload $I_g$, third the hybrid image $O_e$ and fourth column represents the recovered guest image $O_d$.}
\label{figure:ExperimentalResults}
\end{figure*}

\begin{table}
\centering
\resizebox{0.5\textwidth}{!}{
\begin{tabular}{|l|c|c|c|c|c|c|c|c|c|}
\hline
\textit{No.} & \textit{\begin{tabular}[c]{@{}l@{}}Cover\\ Image\end{tabular}} & \textit{\begin{tabular}[c]{@{}l@{}}Payload\\ Image\end{tabular}} & \textit{\begin{tabular}[c]{@{}l@{}}No. of\\ Epochs\end{tabular}} & \textit{\begin{tabular}[c]{@{}l@{}}Avg.\\ bpp\end{tabular}} & \textit{\begin{tabular}[c]{@{}l@{}}Encoder \\ PSNR (db)\end{tabular}} & \textit{\begin{tabular}[c]{@{}l@{}}Decoder \\ PSNR (db)\end{tabular}} & \textit{\begin{tabular}[c]{@{}l@{}}Payload\\  \ \ \ \ \%\end{tabular}} & \textit{\begin{tabular}[c]{@{}l@{}}SSIM \\ Encoder\end{tabular}} & \textit{\begin{tabular}[c]{@{}l@{}}SSIM \\ Decoder\end{tabular}} \\ \hline
1 & \cifar & \mnist & 50 & 7 & 32.9 & 32.0 & 29.1 &0.87 &0.85\\ \hline
2 & \cifar & \cifar & 50 & 8 & 30.9 & 29.9 & 33.3 & 0.98 & 0.96\\ \hline
3 & \imagenet & \imagenet & 50 & 8 & 29.6 & 31.3 & 33.3  & 0.88 & 0.88 \\ \hline
4 & \imagenet & \imagenet & 150 & 8 & 32.9 & 36.6 & 33.3  & 0.96 & 0.96 \\ \hline
\end{tabular}
}
\caption{Comparison of \bpp, \psnr and \ssim values for different runs of our algorithm on different datasets.}
\label{table:comparison}

\centering
\resizebox{0.5\textwidth}{!}{
\begin{tabular}{|l|c|c|c|c|c|c|c|c|c|}
\hline
\textit{No.} & \textit{\begin{tabular}[c]{@{}l@{}}Cover\\ Image\end{tabular}} & \textit{\begin{tabular}[c]{@{}l@{}}Payload\\ Image\end{tabular}} & \textit{\begin{tabular}[c]{@{}l@{}}Avg.\\ bpp\end{tabular}} & \textit{\begin{tabular}[c]{@{}l@{}}Encoder \\ PSNR (db)\end{tabular}} & \textit{\begin{tabular}[c]{@{}l@{}}Decoder \\ PSNR (db)\end{tabular}} & \textit{\begin{tabular}[c]{@{}l@{}}Payload\\  \ \ \ \ \%\end{tabular}} & \begin{tabular}[c]{@{}l@{}}SSIM\\ Encoder\end{tabular} & \begin{tabular}[c]{@{}l@{}}SSIM\\ Decoder\end{tabular} \\ \hline
1 & \pascal & \pascal & 8 & 33.7 & 35.9 & 33.3 & 0.96 & 0.95 \\ \hline
2 & \lfw & \lfw & 8 & 33.7 & 39.9 & 33.3 & 0.95 & 0.96 \\ \hline
3 & \pascal & \lfw & 8 & 33.8  & 37.7 & 33.3 & 0.96  & 0.95 \\ \hline
\end{tabular}
}
\caption{Bpp, \psnr and \ssim values of our \imagenet trained algorithm on different datasets. Note that how our method has been able to generalize across different datasets while maintaining the same payload capacity at similar perceptual scores.}
\label{table:AdditionalExperimentResults}

\end{table}

In this section, we report our experimental settings. We also report quantitative and qualitative results  of our algorithm on a diverse set of publicly available datasets, that is on \imagenet~\cite{deng2009imagenet}, \cifar~\cite{krizhevsky2014cifar}, \mnist~\cite{lecun1998mnist}, \lfw~\cite{huang2008labeled} and \pascal~\cite{everingham2010pascal}. 

We randomly divided each dataset sample images into three datasets: training, validation and testing. All the configurations have been done using validation set and we report the final performance on test set. 

For payload, we randomly select an image from the corresponding dataset and either convert it to gray-scale or just choose a single channel from the RGB channels. For cover, we randomly select an RGB image  from the corresponding dataset.

\def\bpp{bpp\xspace}
\def\psnr{PNSR\xspace}
\def\psnrv#1{#1 db\xspace} 
\def\ssim{SSIM\xspace}

For all experiments we use the same encoder and decoder architecture as explained in \secref{architecture}.  However each input image is zero-centered.  Encoder and decoder weights are randomly initialized using Xavier initialization~\cite{glorot2010understanding}. For learning these weights we use Adam optimizer with a fixed learning rate of 1E-4 and a batch size of 32 where regularization parameter was set to $0.0001$ and $\alpha=\beta=1$.  During each epoch, we disjointly sample images for cover and payload usage from the training set. All the filters in \cnn layers are applied with stride of single pixel and using same padding.

We use Peak Signal to Noise Ratio (PSNR), Structural SIMilarity (SSIM) index and bits per pixel (bpp) to report the perceptual quality of images produced and embedding capacity of our algorithm.


For our initial experiment, we used cover images ($32\times32\times3$) from \cifar while payload images were taken ($28\times28\times1$) from \mnist dataset. For this experiment, we were able to hide approximately 29.1\% payload (\ie 7 \bpp) in our cover images with average \psnr of \psnrv{32.85}  and \psnrv{32.0} for encoder and decoder networks produced images respectively -- \cf Table \ref{table:comparison}. These results show that using our algorithm, we can successfully hide a huge payload with reasonably high \psnr and \ssim values. According to our best of knowledge, no one has been able to report such results on this dataset.

However, \mnist is a relatively simple dataset as majority of pixels in each image belong to plain background (white color) class. Thus, we conducted another experiment on \cifar dataset -- \cifar being dataset of natural classes contains much larger variation in image foreground and background regions -- with identical experimental settings.

In this experiment, both cover ($32\times32\times3$) and payload images ($32\times32\times1$) were randomly and disjointly sampled from \cifar training batch. In this experiment we were able to hide a payload of 33.3\% (\ie 8 bpp) in our cover images with average \psnr of \psnrv{30.9} and \psnrv{29.9} for encoder and decoder networks produced images respectively. 

From our these experiments, we can conclude that our proposed algorithm is extremely generic and one can, using the same architecture,  reliably guarantee huge payloads and acceptable \psnr values for complex images as well -- \cf \tabref{comparison}. For both these experiments we ran our algorithm for 50 epochs.

To further consolidate our findings and to evaluate our algorithm's embedding capacity and reconstruction performance on images of large size, we designed another experiment using \imagenet dataset.  A subset of 8,000 images was randomly chosen from one million images. These selected images were then divided into two disjoints sets: training (6,000 images) and testing (2,000 images) -- no validation set was used here since we reuse the earlier experiments settings. To allow uniform sized images as cover and payload all of these images were then resized to $300\times300$ pixels. 
For our initial version of this experiment and to ensure a fair comparison with other results, we first ran our algorithm for 50 epochs.

For randomly sampled cover ($300\times300\times3$) and guest images ($300\times300\times1$) from our \imagenet test dataset, we were able to hide a payload of 33.3\% (\ie 8 bpp) in our cover images with average \psnr of \psnrv{29.6} and \psnrv{31.3} for encoder and decoder networks produced images respectively. As we were able to hide high payload for similar \psnr values to earlier experiments for this complex dataset as well, so we further explored different experimental settings. 

Our final model on \imagenet was trained for 150 epochs further improving the \psnr values for encoder and decoder to \psnrv{32.92} (\ssim=0.96) and \psnrv{36.58} (\ssim=0.96) respectively from \psnrv{29.6} and \psnrv{31.3} while maintaining similar payload capacity of 33.3\% (on average 8 \bpp) -- \cf \tabref{comparison}.

To further evaluate the generalization capacity of our algorithm, we ran the \imagenet trained algorithm on sample of 1,000 unseen images from \pascal~\cite{everingham2010pascal} and Labelled Faces in Wild (\lfw) ~\cite{huang2008labeled} datasets. \tabref{AdditionalExperimentResults} shows the results of our this experiment. Here, even though our algorithm is trained on different dataset, it is still being able to achieve high payload capacity at high \psnr and \ssim values which shows the generalization capabilities of our proposed algorithm.

\figref{ExperimentalResults} shows a sample of result images from \lfw, \pascal and \imagenet datasets. Here once again we can verify using qualitative analysis that our method is being able to conceal and recover unseen complex payload images. 

Therefore, given this quantitative and qualitative analysis, we can conclude that our algorithm is generic and robust to complex backgrounds and variations in objects appearance, thus can be reliably used for image steganography.  


\section{Conclusions}\label{sec:conclusion}
In this paper, we have presented a novel CNN based encoder-decoder architecture for image steganography.  In comparison to earlier methods, which only consider binary representation as payload our algorithm directly takes an image as payload and uses a pair of encoder-decoder networks to embed and robustly recover it from the cover image. According to our best of knowledge, no such earlier work exists and we are the first one to introduce this method for image-in-image hiding using deep neural networks.  We have performed extensive experiments and empirically proven the superiority of our proposed method by showing excellent results with strong payload capacity on a wide range of wild-image datasets. 
\bibliographystyle{IEEEbib}
\bibliography{refs}
\end{document}